\documentclass{emulateapj} 

\usepackage{ifthen}
\newboolean{color}

\setboolean{color}{true}

\slugcomment{\\Submitted ApJ}

\usepackage{amsmath}
\usepackage{graphicx}
\usepackage{float}
\usepackage{amssymb,amsmath}
\usepackage{epsfig,floatflt}
\usepackage{subfigure}
\usepackage{apjfonts}

\shorttitle{Phase lag of Cepheids}
\shortauthors{Szab\'o, Bartee, Buchler}

\begin{document}

\newcommand \thi{\thinspace}
\newcommand \noi{\noindent}

\newcommand \eg{{{\it e.g.},\ }}
\newcommand \etal{{\it et al.\ }}
\newcommand \etc{{\it etc.\ }}
\newcommand \cf{{\it cf.\ }}
\newcommand \ia{{{\it inter alia},\ }}
\newcommand \ie{{{\it i.e.},\ }}
\newcommand \via{{\it via\ }}
\newcommand \viz{{\it viz.\ }}
\newcommand \vs{{\it vs.\ }}

\newcommand \Teff{{$T_{\rm {ef\!f}} $}}
\newcommand \teff{{T_{\rm {ef\!f}} }}
\newcommand \Lo{{$L_\odot $}}
\newcommand \Mo{{$M_\odot $}}
\newcommand \Log{{\rm Log}\thi}
\newcommand \approxgt{\,\raise2pt \hbox{$>$}\kern-8pt\lower2.pt\hbox{$\sim$}\,}
\newcommand \approxlt{\,\raise2pt \hbox{$<$}\kern-8pt\lower2.pt\hbox{$\sim$}\,}
\newcommand\dotd{\hbox{$.\!\!^{\rm d}$}}
\newcommand\dotm{\hbox{$.\!\!^{\rm m}$}}
\newcommand\dd{{$^d$}}
\newcommand\ML{{$M$-$L$\ }}
\newcommand\aanda{A\&A}
\newcommand\RV{$V_r$}
\newcommand\DP{{$\Delta(\Delta\Phi_1)$}}
\newcommand\DDP{{$|\Delta(\Delta\Phi_1)|$}}

\title{The Cepheid Phase Lag Revisited}

\author{R\'obert Szab\'o\altaffilmark{1}, J. Robert Buchler and Justin Bartee}
\affil{Department of Physics, University of Florida,
    Gainesville, FL 32611-8440}
\email{rszabo@konkoly.hu, buchler@phys.ufl.edu}

\altaffiltext{1}{On leave from Konkoly Observatory, Budapest, Hungary}

\begin{abstract}

We compute the phase lags between the radial velocity curves and the
light curves $\Delta \Phi_1= \phi^{V_r}_1 - \phi^{mag}_1$ for classical Cepheid
model sequences both in the linear and the nonlinear regimes.  The nonlinear
phase lags generally fall below the linear ones except for high period models
where they lie above, and of course for low pulsation amplitudes where the
two merge.  The calculated phase lags show good agreement with the available
observational data of normal amplitude Galactic Cepheids.  The metallicity has
but a moderate effect on the phase lag, while the mass-luminosity relation and
the parameters of the turbulent convective model (time-dependent mixing length)
mainly influence the modal selection and the period, which is then reflected in
the Period -- $\Delta \Phi_1$ diagram.  We discuss the potential application of
this observable as a discriminant for pulsation modes and as a test for
ultra-low amplitudes (ULA) pulsation.

 \end{abstract}


\section{Introduction}

The relative phase between the luminosity and the velocity curves in classical
pulsating variables was a puzzle in the early days of variable star modeling.
Because, overall, the pulsations are only weakly nonadiabatic it was expected
that the maximum brightness should occur at maximum compression, \ie minimum
radius, whereas in reality it is observed to occur close to maximum velocity.
(To avoid confusion we note that in this paper we use astro-centric
velocities, $u = dR/dt$.)  
The luminosity thus has a $\sim 90 \deg $ phase lag compared to
the one expected adiabatically.

As it became possible to make accurate linear and nonlinear calculations of the
whole envelope, including in particular the outer neutral hydrogen region,
agreement between modeling and observation was achieved.  However, it was
\cite{cas68} who first provided a qualitative {\it physical} understanding of
this phenomenon.  He pointed out that during the compression phase, for example,
the hydrogen partial ionization front moves outward with respect to the fluid,
and energy is removed from the heat flow to ionize matter.  
Because of this temporary storage
of energy - as in the charging of a capacitor - the H acts acts like a low-pass
filter, causing a phase lag of close to 90 degrees.  However, the exact value
of the phase lag depends sensitively on details of the stellar model, and can
only be obtained by detailed simulations.  An understanding of the phase lag
involves physics both in the linear and the nonlinear regimes.  This
characteristic makes it an ideal benchmark to test existing hydrocodes against
observational constraints.

Cepheids are known to have humps or bumps on the light curves, especially near
maximum light.  Rather than define the phases with respect to maximum light,
or some other chosen reference point, it is therefore desirable to define them
in terms of a Fourier sum:
\begin{equation}
f(t) = \sum_{k=1}^n A_k\thi \sin (k\thi \omega \thi t + \phi_k)
\label{eq_ffit}
\end{equation}
\noi The phase lag $\Delta \Phi_1$ =($\phi^{V_r}_1 - \phi^{mag}_1$) is thus
defined as the difference between the  phases of the fundamental components
of the Fourier fits of the radial velocity (\RV) and magnitude curves.  The two
time sequences are of course reduced to the same time origin (epoch).

Good Galactic Cepheid light curve data have been available for a long time,
while radial velocity data are more recent.  \cite{ogl00} have published the
phase lags for a set of fundamental (F) and first overtone (O1) Cepheids and have
found that for the F Cepheids $\Delta \Phi_1$ is largely independent of
pulsational period, with an average value of --0.28, but that for
the O1 Cepheids it decreases with period from $-0.24$ to $-0.71$.

In this paper we describe the results of an extended survey of the linear and
the nonlinear (full amplitude) Cepheid pulsation models, and we revisit the
phase lag problem.  In our computations we make use of the state-of-the-art
Florida-Budapest hydrocode \citep{ykb, kbby98, kbsc02} which includes a model
for turbulent convection.

Our goal is to compare the modeling results with a turbulent convective
hydrocode for the phase lag with the available observational data.  The last
similar systematic phase lag investigation was carried out by \cite{sd83} using
early radiative hydrocodes and opacities.  Furthermore, the limited number of
their models, and the quality of the observations that were available at the
time justify a new look at the problem.

\begin{figure}
\epsscale{1.13}
\ifthenelse{\boolean{color}}
{\plotone{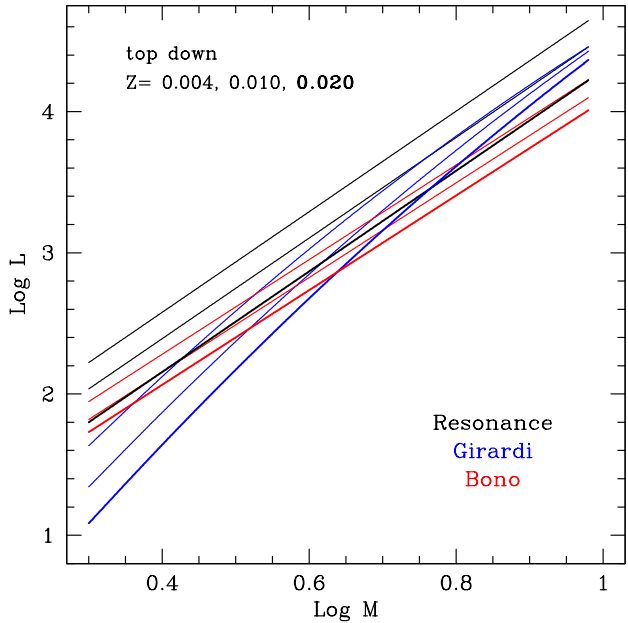}}
{\plotone{fig1.eps}}

\caption{The Resonance, Girardi and Bono mass-luminosity relations for
Galactic, LMC and SMC metallicity.
}

\label{fig:masslum}
\end{figure}


\section{Linear and Nonlinear Cepheid Modeling}

With the usual convenient, but excellent assumptions of a chemically
homogeneous envelope and a rigid, nonpulsating core, we can then specify a
Cepheid envelope by three parameters for a given composition: effective
temperature (\Teff), luminosity ($L$) and stellar mass ($M$).  
The composition is usually specified by the H content $X$, the
metallicity $Z$ and the He content $Y$=1--$X$--$Z$.

Cepheids evolve on tracks that are provided by stellar evolution calculations
$L=L(t; M)$ and $\teff=\teff(t; M)$,
and the vast majority of Cepheids are expected to be found on the second,
blueward penetration of the instability strip ({\bf IS}).  Since the Cepheid 
tracks are more or less
horizontal (fixed $L$) it is customary and expedient (but not necessary) 
to identify these tracks by an \ML relation, $L=L(M\thi ;\thi X,Z)$.  
This reduces the number of stellar
parameters by one, and we can thus make one-parameter sequences of models for a
given mass $M$ and composition, with \Teff\ as the single parameter.  The selected \ML
relation however affects the whole picture from modal selection, to mass-period
relation, to phase lags.

The \ML  relations
are usually obtained from evolutionary calculations.  The problems with
evolutionary tracks producing Cepheids and the difficulty of deriving
mass-luminosity relations are well known and are addressed in \citet{bs07}.  
In this paper we make use of three different \ML relations.  
Our first \ML
relation is a fit to the evolutionary tracks with a nonzero overshooting
parameter of \cite{girardi}.  
Second, \citet{bono} give an \ML relation that is a fit
to their evolutionary tracks without overshooting.  Third, one can also bypass
evolution calculations and combine observational data with pulsational theory.
This has been done for the SMC and the LMC by \cite{bbk} who derived stellar
masses from the observed magnitudes, colors and periods, but reddening errors
prevented the extraction of a very tight \ML relation.  An alternative is to
pinpoint the $P_2:P_0=1:2$ resonance center from the behavior of the Fourier
phase $\phi_{21}$ as a function of period, and then use linear modeling to
match the resonance condition and thus tie down a 'Resonance' \ML relation,
which is assumed to be linear: $\Log\thi L = a + 3.56\thi \Log\thi M$, where $a$
is 0.7328, 0.96864, 1.1552 for $Z$ = 0.020, 0.010 and 0.004, respectively.  We
refer to it as the {\it Resonance} \ML relation.  (In this paper we label the
modal periods with numeral subscripts, \ie 0 for F, 1 for O1, etc.)

The three \ML relations are displayed in Fig.~\ref{fig:masslum}.  For Z=0.020
the Resonance \ML relation runs close to the Bono relation in the low mass
regime, while the Girardi relation predicts much lower luminosity.  In the high
mass range ($M > 5.8$\Mo) the Resonance \ML relation runs between the Girardi
and Bono formulae for this metallicity.  This characteristic has led us to
choose the Resonance \ML relation as the reference throughout this paper.  The
sensitivity to the form of the \ML relation will be discussed in
Sec.~\ref{sec_ML}.

The mass range of our survey extends from a lower $M$ of 2.0\Mo\ to an upper
$M$ that we vary with the metallicity and observational constraints, \ie the
observed longest period pulsation, and is 9.5\Mo, 7.5\Mo\ and 6.5\Mo\ for
Galactic, LMC and SMC variables, respectively.  The mass increment is 0.5 \Mo\
between the sequences.  The luminosity is obtained from the \ML relation.  The
temperature range covers the whole classical IS, and
subsequent models are computed 100\thi K apart.

In our calculations we employ the Florida-Budapest hydrocode \citep{ykb,
kbby98, kbsc02} to produce equilibrium Cepheid models, to perform their linear
stability analysis and to compute full amplitude pulsations.  This code
incorporates a time-dependent mixing length model for turbulent convection
\citep{ykb}.  The G91 OPAL \citep{iglesias96} with solar element abundances
\citep{grevesse} and \cite{af94} opacities are used.  The equation of state and
the opacity are given in terms of the usual composition variables, $X$, $Y$ and
$Z$.

\begin{table}
\caption{\small Turbulent convection parameters}
\vspace{-5mm}
\begin{center}
\medskip
\begin{tabular}{c c c c c c c c c}
\hline\hline
    \noalign{\smallskip}
& $\alpha_d$ & $\alpha_c$& $\alpha_s$ & $\alpha_{\nu}$& $\alpha_t$ & $\alpha_r$ & $\alpha_p$ 
& $\alpha_{\lambda}$ \\
    \noalign{\smallskip} 
    \hline   
     \noalign{\smallskip}
A  & 2.177 & 0.4 & 0.4330 &  0.12 & 0.001 & 0.4 & 0.0 & 1.5  \\  \noalign{\smallskip}
\hline
B & 2.177 &  0.4082  & 0.4082 & 0.25 & 0.0 &  0.0 & 0.0 &  1.5 \\  \noalign{\smallskip}
\hline
\end{tabular}
\label{tab_alphae}
\end{center}
\end{table}

The values of the turbulent convective ($\alpha$) parameters in the code affect
most of the observables of Cepheid models, such as the width of the IS, the
shape of the light and radial velocity curves, the amplitudes, the locations of
the resonances, the maximum overtone period, just to name a few.  It has been
found that very different parameter sets can give surprisingly similar results.
But a global calibration of the turbulent convective parameter set has not been
performed yet, because of the huge computational task.  Therefore we restrict
ourselves to only two sets, which are presented in Table~\ref{tab_alphae}.
The 8 parameters are dimensionless, of order of unity, but 
they obey a scaling invariance \citep{kbsc02}.  For a comparison to other
mixing length models see \cite{fbk00}.

The linear code provides the radial displacement, velocity, temperature and
turbulent energy eigenvectors for all the radial pulsation modes at any point,
and in particular in the photosphere.  From these we can compute the luminosity
and magnitude eigenvectors, and then the phase lags $\Delta \Phi_1$.  In
this paper we are interested in periodic F, O1 and O2 pulsations (limit
cycles).

The equilibrium models are initiated with a velocity kick in one of the
eigenmodes and the desired full amplitude (limit cycle) pulsations are computed
with our hydrocode.  The regime where the F and O1 modes are simultaneously
linearly unstable is well known to split into two types, depending on the
stellar parameters, namely a double-mode regime, where both modes achieve full
amplitude, and a hysteretic regime -- the 'either--or' regime in the pulsation
jargon -- where F and O1 limit cycles coexist, and the initial kick determines
which of the two limit cycles is attained.  A similar split occurs for the O1
and O2 modes. (In some narrow regions the situation can be even more
complicated, \eg \cite{kbsc02}).  The double-mode pulsations (also called beat
pulsations) are not of interest to us here, but are taken up in a subsequent
paper.  The desired phase lags $\Delta \Phi_1$
are obtained from the magnitude and \RV\ curves of the computed limit cycles
through an 8th order Fourier fit (Eq.~\ref{eq_ffit}).

\begin{figure}
\epsscale{1.1}
\ifthenelse{\boolean{color}}
{\plotone{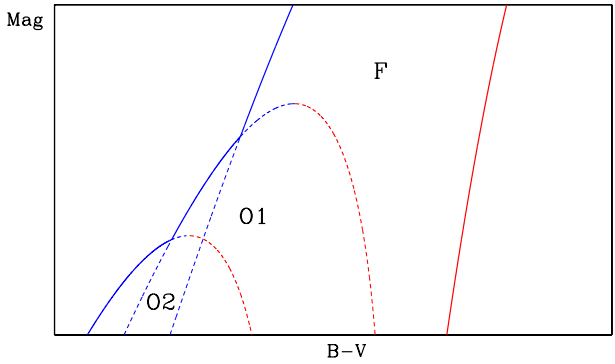}}
{\plotone{fig2.eps}}

\caption{Schematic HR diagram showing the {\it linear} instability strips of
the fundamental (F), first (O1) and second overtone (O2) modes.  At the solid
lines the linear and nonlinear edges coincide.}

\label{fig:is}
\end{figure}

\section{Modal Selection -- The Bifurcations}

First we recall the general topography of the {\it linear} IS shown
schematically in an HR diagram on Fig.~\ref{fig:is}.  In the domain of astrophysical
parameters that are relevant for Cepheids, only the F, O1 and O2 modes are
unstable.  As to the definitions of linear and nonlinear edges we refer to the 
literature, \eg \citep{stellingwerf75,bk86,sb07} and in particular Figs.~4
and 5 in \cite{fbk00} whose notation we will use here, namely {\bf L} for
linear and {\bf N} for nonlinear, with {\bf BE} for the blue edge
and {\bf RE} for the red edge.  Fig.~\ref{fig:is} shows that the region of
linear instability of the F mode extends to very high luminosities.  That of
the O1 mode resembles a sugar cone, thus with a maximum luminosity, which
overlaps with the F IS.  Similarly the O2 IS sugar cone, with a yet lower
maximum luminosity, overlaps with both the O1 and F ISs.  Along the solid lines
the linear and nonlinear edges coincide, and the pulsation amplitudes vanish
there \thi (See \cite{bk02} for a more thorough discussion of how the behavior
at the edges of the IS is affected by stellar evolution).  The dashed parts of
the linear instability boundaries are not real edges because an evolving star
that is pulsating in a given mode undergoes a switch (bifurcation) to another
type of pulsation before it can reach the dashed line corresponding to the
original pulsation mode.  The substructure in the topography of the IS will be
discussed in a paper on Beat Cepheids \citep{sb07}.

For the purposes of this paper we note that because the star pulsates with
infinitesimal amplitude near the solid edges, the linear and nonlinear phase
lags therefore approach each other.  All other mode switching occurs
with finite amplitude and the linear and nonlinear $\Delta\Phi_1$ differ
therefore at the bifurcation.  This behavior will clearly be visible in the
subsequent figures.  Thus for example, a star on a blueward horizontal track at
high luminosity in Fig.~\ref{fig:is} starts to pulsate with infinitesimal
amplitude when it crosses the FRE (the solid line on the right) and stays in
this mode until it reaches the FBE (solid line) on the left where its pulsation
amplitude goes to zero.  The linear and nonlinear $\Delta\Phi_1$ should thus be
equal to each other both at the FRE and at the FBE.  In contrast, a star on a
blueward track at mid-magnitude in Fig.~\ref{fig:is} similarly starts to
pulsate when it crosses the FRE (the solid line on the right), but it
bifurcates to some other pulsational state before it reaches the LFBE (the
dotted line corresponding to the F mode).  The amplitude of the pulsation does
not go to zero at the switching which has as a consequence that the linear and
nonlinear $\Delta\Phi_1$ are different.  Finally, we note that because of the
\ML relation, the ordinate in the schematic Fig.~\ref{fig:is} could also
represent the stellar mass $M$.

In the two hysteretic regimes, where the pulsation is periodic, it can be in
either the F or the O1 mode, or it can be in either the O1 or O2 mode,
respectively.  In this regime we include the corresponding model in both the F
and the O1 set of pulsators.  The double-mode pulsations of two types that have
been encountered, namely F\&O1 and O1\&O2, are ignored here because of their
multi-periodic nature.  They will be discussed in a separate paper
\citep{sb07}.  The temperature range of the double-mode region is typically
less then $100K$, therefore the exclusion of these models has little effect on
the results of this paper.

\begin{figure*}
\epsscale{1.15}
\ifthenelse{\boolean{color}}
{\plottwo{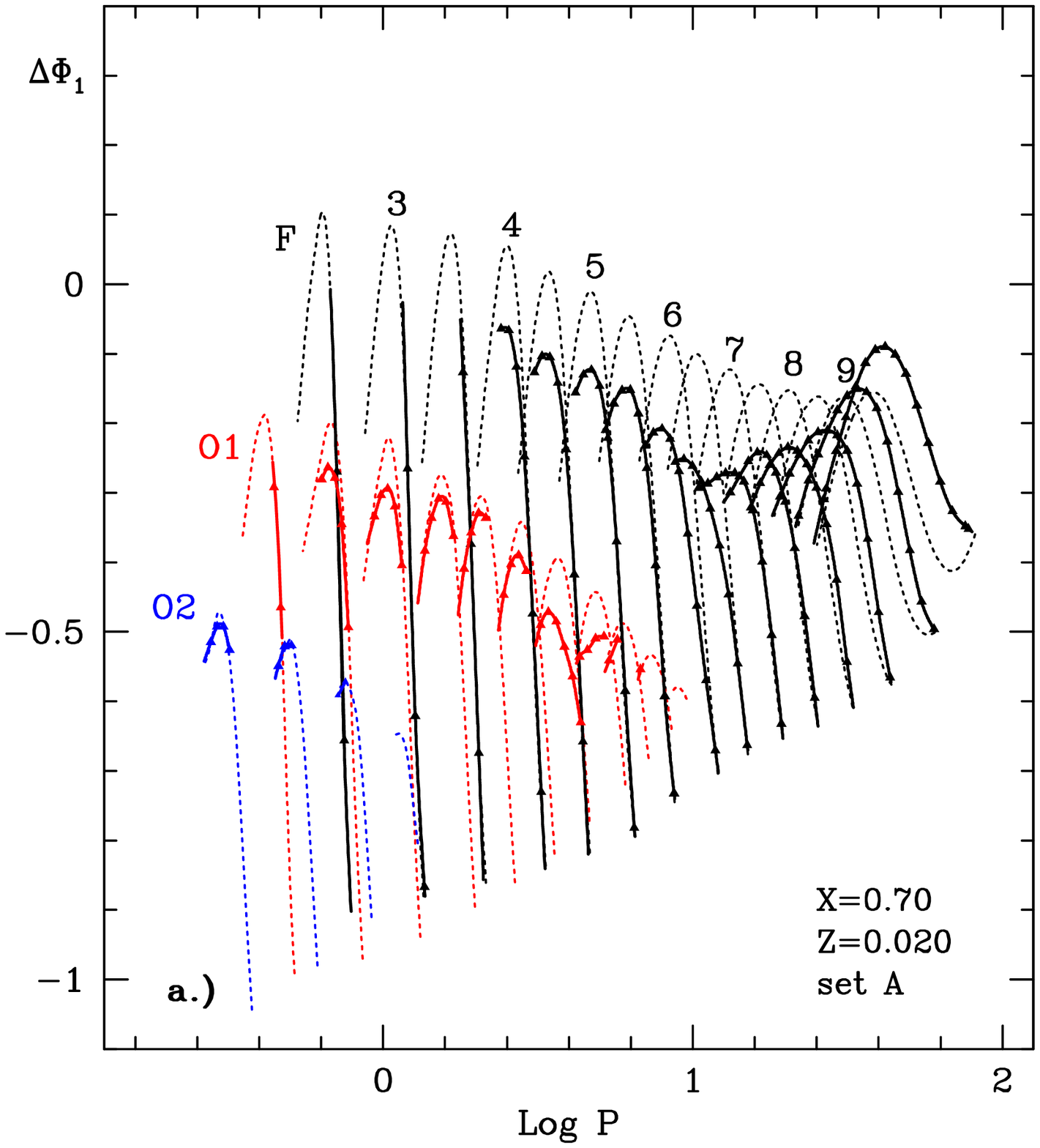}{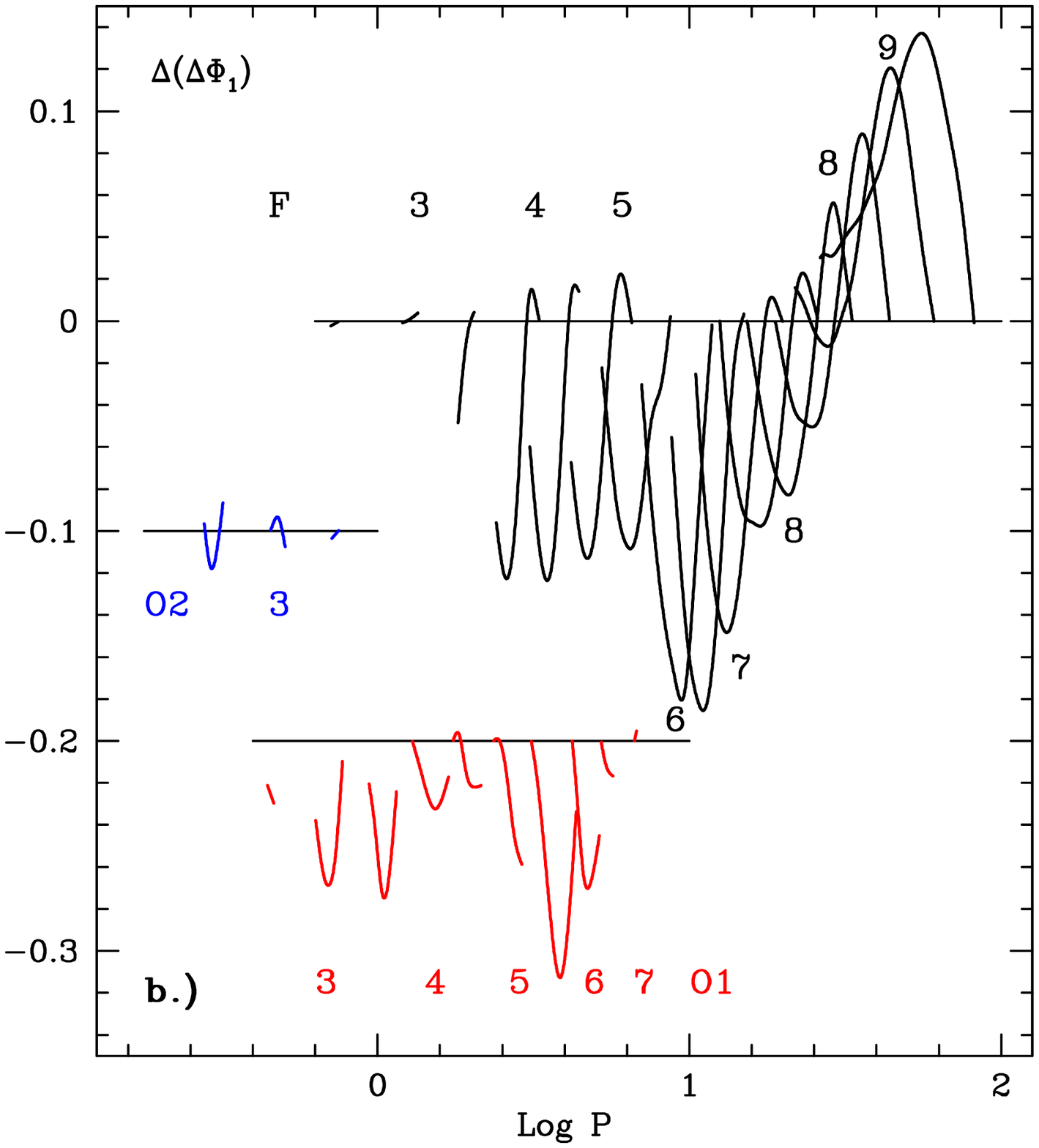}}
{\plottwo{fig3a.eps}{fig3b.eps}}

\caption{{\bf a.)} Linear and nonlinear phase lags of sequences of 
equal mass Galactic Cepheid models. The mass labels appear on top.
The dashed lines denote the linear phase lags of linearly unstable models, 
the thick lines denote the nonlinear (limit cycle) phase lags.  
The triangles along the model sequences denote  100K intervals in \Teff.  
The Resonance \ML relation was used along with the $\alpha$ parameter set~A \thi 
(Table~\ref{tab_alphae}). {\bf b.)} The difference between the nonlinear 
and linear phase lags (\DP) for the same models, 
as a function of the pulsational period. The O1 and O2 models are shifted 
vertically by $-0.2$ and $-0.1$, respectively for clarity.
See color version of the figures in the online edition.  }

\label{fig:linNL}
\end{figure*}

\section{Linear and Nonlinear Phase Lags}

\subsection{Results}

The general picture of the theoretical linear and nonlinear phase lags is
displayed in Fig.~\ref{fig:linNL}a.  These results are obtained by using the
Resonance \ML relation, the turbulent convective parameter set~A
(Table~\ref{tab_alphae}), and a Galactic metallicity (Z=0.020).  Below we
investigate how each of these three choices affects the results.  The phase
lags are plotted as a function of the logarithm of the period of the excited
mode\thi\thi (The models in the 'either F or O1' regime appear twice, once under
$P_0$ and once under $P_1$).  Our models form one-parameter sequences at a
given $M$, and thus $L=L(M)$, with \Teff\ as the variable parameter.  The
masses of the sequences run from 2.5\Mo\ to 9.5\Mo, with a 0.5\Mo\ increment,
and are labeled by the small (integer) numbers.  The F, O1 and O2 groups start
with the same masses on the left.  The dotted lines depict the linear phase
lags of the linearly unstable models, while the thick lines correspond to the
nonlinear phase lags, \ie those of the limit cycles.  (Some inter/extrapolation
has been done near the edges.)

In Fig.~\ref{fig:linNL}a one can identify the three separate clusters that
correspond to the F, O1 and O2 radial pulsation modes of the sequences of
models.  For high masses, the nonlinear sequences follow the linear ones of the
same mass.  At a certain mass O1 pulsation appears, thus the nonlinear F curves
stop before reaching the FBE, and the sequence continues in the corresponding
O1 curve.  In the 'either-or' regions, the same model contributes to two
pulsational modes.  The same thing happens where the sequences change from O1
to O2 pulsation at even lower mass.  Therefore the F and O1 sequences seem to
be increasingly incomplete toward the short periods at low mass. At the lowest
mass regime, the models reach the blue edge pulsating in the O2 mode.

As expected, the F mode pulsations cover the broadest range in period and in
mass, as well as in phase lag.  We remark in advance that when comparing
the theoretical values of the phase lags to observations the most relevant part
of these arches are the broad vicinity of their maxima, because the long 'legs'
on the high period (right) side of the sequences are associated with low
pulsational amplitudes, near the FRE.  We will return to this issue.
Along the sequences the dots mark 100\thi K intervals which allows one to see
that the models are not equidistant along the $\Delta\Phi_1$ in
Fig.~\ref{fig:linNL}, but that they crowd toward the center of the IS and are
increasingly spaced out toward the FRE.  The FRE is the lower envelope of the F
sequences on the figure, corresponding to infinitesimal pulsational amplitude.
The spacing is increasing toward the low mass models, too.
The maximum F phase lags of the models thus  lie close to
zero, while the bulk of our models are in the --0.1 to --0.4 range.

The phase lags for the O1 and the O2 pulsations of the sequences are located
lower on the diagram to the left.  With the chosen set~A of $\alpha$ parameters
the maximum O1 period is $\sim 7$\dd, and the maximum O2 period is $\sim1$\dd.
The Cepheids pulsating with the maximum O1 period are close to the LOBE,
and therefore low amplitudes are expected to hamper their discovery (see
\cite{fbk00} for a similar comment).  The same argument applies for the
maximum O2 period, although no Galactic O2 Cepheids have been found to date.

We need to point out that this maximum O1 period is sensitive to
the turbulent convective parameters that we use.  This will be discussed in
Sec.~\ref{alphas} and a thorough calibration will be attempted in
\cite{sb07}.  We note however, that our phase lag results do not seem very
affected by changes in $\alpha$s as long as we use a parameter combination that
reproduces the maximum O1 period.

\begin{figure*}
\epsscale{1.2}
\ifthenelse{\boolean{color}}
{\plotone{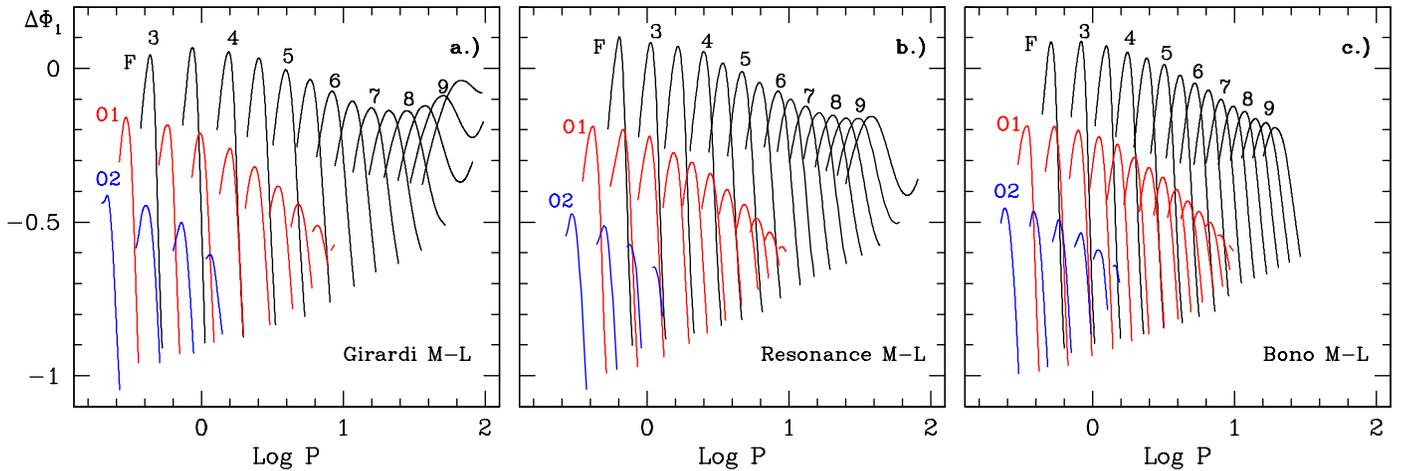}}
{\plotone{fig4.eps}}

\caption{Linear phase lags for Galactic metallicity (Z=0.020) with three
\ML relations: {\bf a.)} Girardi, {\bf b.)} the Resonance \ML
relation, and {\bf c.)} Bono, for the $\alpha$ parameter set~A. }

\label{fig:MLrel}
\end{figure*}

\subsection{Resonances} \label{sec_res}

It is well known that internal resonances cause structure in the Fourier
coefficients \vs period or some other parameter, (\eg \cite{buchlermito}).
Thus the light curves of RR Lyrae who have no low-order internal resonances
exhibit very smooth Fourier decomposition coefficients ($\phi_{21}, R_{21}$,
etc), whereas the Cepheids which span a much wider range of $M$, $L$ and \Teff\
encounter a number of such resonances during their evolution through the IS (see
for example Fig.~4 in \cite{buchler97}).  We will see that these resonance
characteristics play a role in this study.
This may come as a surprise because one might have thought that resonances
affect for light curves and radial velocity curves and that their effect would
balance out in $\Delta\Phi_1 = \phi_1^{V_r} - \phi_1^{mag}$.  This is however
not the case because the resonance induced distortions  are different for light
and for radial velocity.

As we mentioned earlier, the differences between the nonlinear and linear phase
lags, \DP, vanish at the IS edges.  This is depicted on 
Fig.~\ref{fig:linNL}a, especially for the F and O1 sequences.  The \DDP\ are
very small for the O2 pulsators which have very low amplitudes.  Generally, one
expects that the larger the amplitude the larger the difference between the
linear and nonlinear quantities such as the phase lag.  This simple argument
may break down where there are resonances which distort the light and
radial velocity curves.

The largest \DDP\ are found around $P_0$ = 10\dd\ for the F pulsators, but it
is not immediately clear whether this is related to the well-known $P_2:P_0 =
1:2$ resonance, or whether it is just a broad minimum occurring there by
coincidence, as part of a general trend running through the whole mass range.
To resolve the issue, we display the \DP\ values on 
Fig.~\ref{fig:linNL}b.  The O1 and O2 sequences are shifted vertically for
clarity. The \DP\ are positive for the highest mass values, and tend to be 
negative for lower masses.  For a given mass sequence its absolute value is 
largest toward the middle of the instability strip. We point out that this figure 
is very similar for the \DP\ for LMC and SMC metallicities that are discussed in 
Sec.~\ref{met_eff}. 

The \DP\ tend to be more negative in the vicinity of the resonance, and they do
not extend to the positive region.  On the low period side \DP\ changes
abruptly due to the distortion of the light and velocity curves that is caused
by the resonance.  Based on these findings, we conclude, that the $P_2:P_0 =
1:2$ resonance has a noticeable effect on the phase lags. Namely, an additional dip
appears on the upper envelope of the arched nonlinear phase lags which is very
pronounced around the resonance center.  The Log P -- \DP\ diagram shows, that
the resonance related distortion plays an equally important role in shaping the
phase lag curves as the amplitudes themselves.  The same resonance behavior
will be noticed in the LMC and the SMC metallicity models around the
apposite resonance periods confirming our argument.

Similarly, we believe that the surprisingly large $|\Delta\Phi_1|$ that are
found for the O1 modes of the 5.5 \Mo\ sequence are associated with the
$P_4:P_1 = 1:2$ resonance near an overtone period $P_1 = 4\dotd 2$ period.
This resonance causes the 'either or region' to broaden to a width of several
hundred K \citep{sb07}, which is in turn reflected in the much longer O1 phase
lag curve.  The much shorter 6\Mo\ curve is still affected by the resonance, as
its strange shape and the large \DP\ indicate.

The effect of these resonances can also be witnessed on the amplitude behavior 
which is discussed further in Sec.~\ref{sec_amplitudes}.

\begin{figure*}
\epsscale{1.15}
\ifthenelse{\boolean{color}}
{\plotone{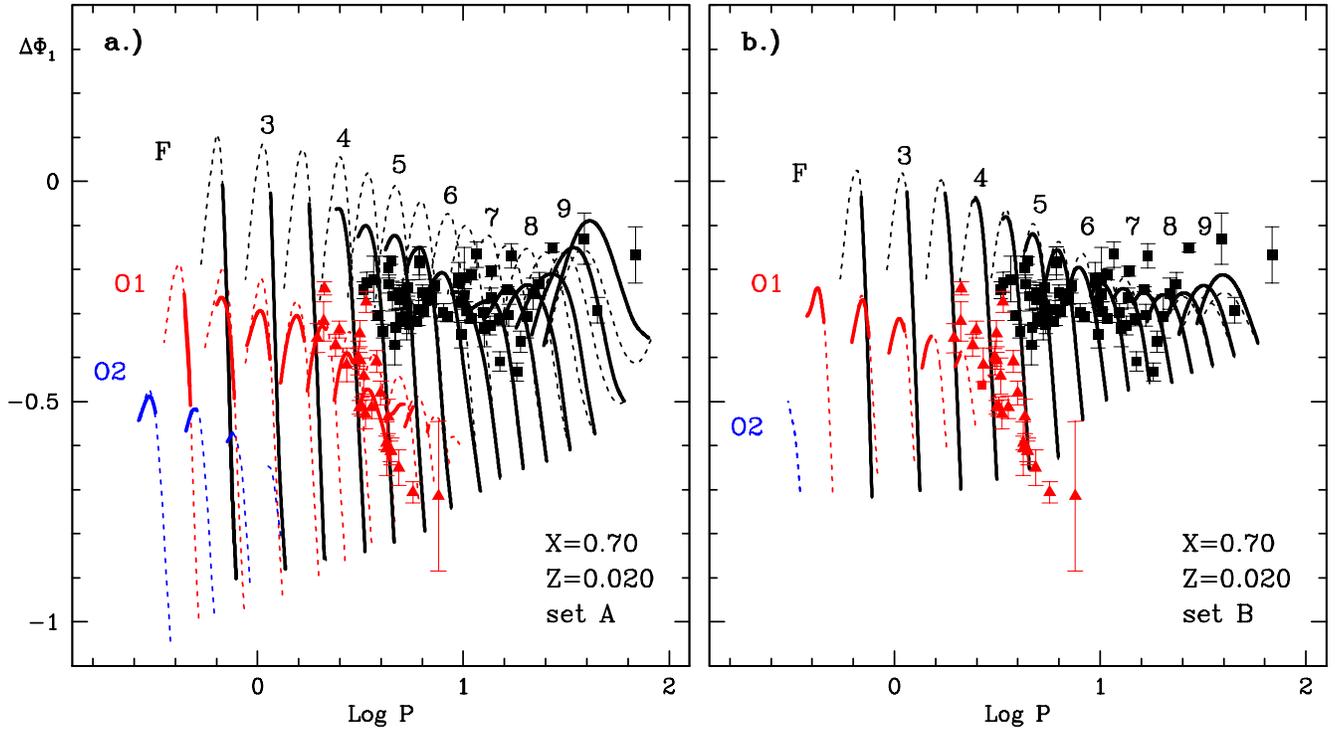}}
{\plotone{fig5.eps}}

\caption{Linear and nonlinear {\it Galactic} phase lags computed with two 
different $\alpha$ parameter combinations a.) set~A and b.) set~B. and using 
the Resonance \ML relation.  Dashed lines: linearly unstable models, thick lines: 
nonlinear (limit cycle) phase lags.  The observed phase lags with error
bars are overplotted for F (squares) and O1 (triangles) Cepheids \citep{ogl00, mo00}.
See color version in the online edition.  }

\label{fig:obs_alpha}
\end{figure*}
\vskip 10pt

\begin{figure*}
\epsscale{1.15}
\ifthenelse{\boolean{color}}
{\plotone{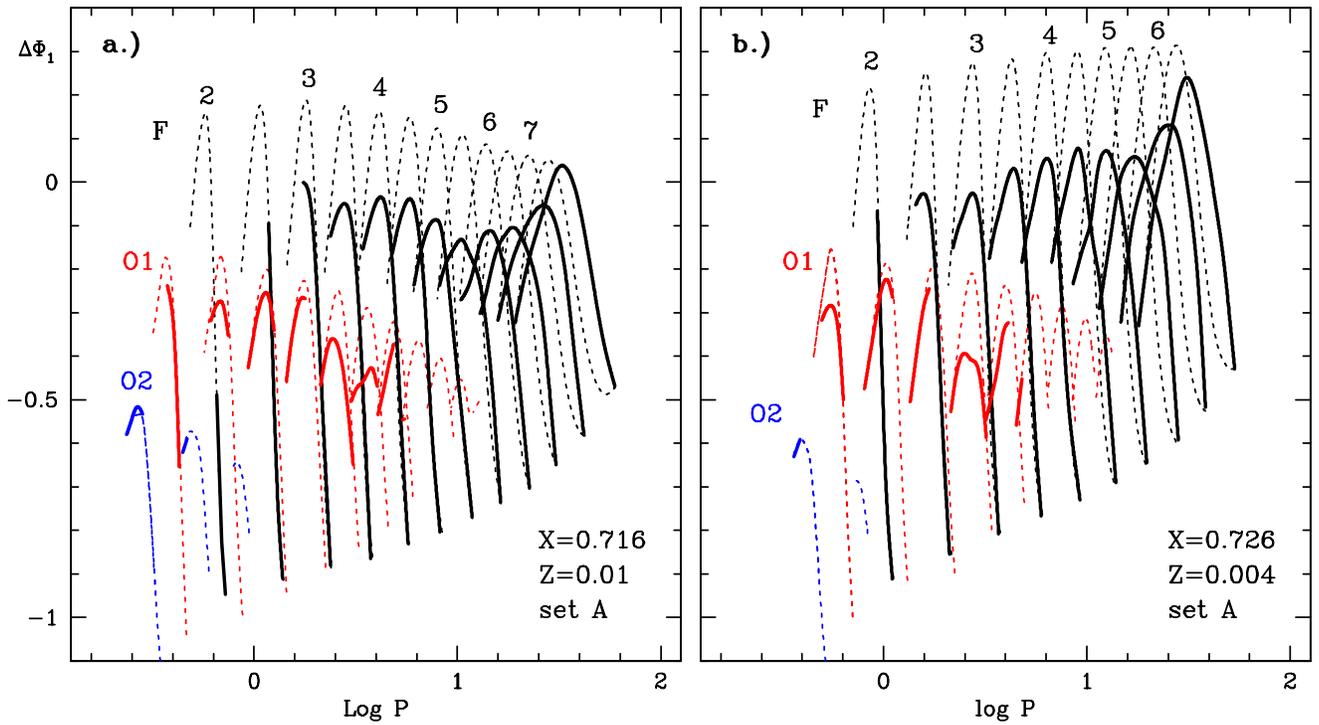}}
{\plotone{fig6.eps}}

\caption{The effect of metallicity on the phase lags. LMC and SMC models are shown 
on panel a.) and b.), respectively. Dashed lines: linearly unstable models, 
thick lines: nonlinear (limit cycle) phase lags. The results are obtained with the 
Resonance \ML relation and the $\alpha$ parameter set~A.}

\label{fig:metal}
\end{figure*}

\subsection{Effects of the \ML relation} \label{sec_ML}

Fig.~\ref{fig:MLrel} compares the linear $\Delta\Phi_1$ for the three \ML
relations that we have described in \S2.  In order to avoid overcrowding we
plot only the linear phases lags.  The overall difference between the
nonlinear and linear lags are found to be similar to what we have seen in
Fig.~\ref{fig:linNL}a for the Resonance {\ML.}

One immediately notices that the effect of the three \ML relations is most
prominent in the high mass regime, where the highest periods are quite
different.  The Bono \ML relation has lower luminosity than the other two, and
it would require a much higher mass ($>10$\Mo) to reach the 100\dd\ fundamental
period limit to which the 9.5\Mo\ sequence rises for the Resonance \ML.  On the
low-mass and low luminosity end the 1\dotd 0 \thi F period occurs for $3.50$,
$3.00$ and $3.25$\Mo\ Cepheids, for the Girardi, Resonance and Bono \ML
relations, respectively, as one would expect from Fig.~\ref{fig:masslum}.

With the Girardi relation the $\Delta\Phi_1$ values of the models 
turn upward (especially on the low-temperature (right) side).  The
same trend is visible with the Resonance \ML relation but does not occur with
the Bono one, again consistently with the behavior of the \ML relations.

We conclude that the effect of the \ML relations on the $\Delta\Phi_1 -
\log P$ diagram occurs indirectly through a modification of the
mass-period relation.

Generally the observed phase lags, which are shown in
Fig.~\ref{fig:MLrel}, and the mass-period distribution show good agreement with
the models obtained by the Girardi and the intermediate \ML relations, but
disagree with the Bono relation at high mass.

\subsection{Effects of the $\alpha$ Parameters} \label{alphas}

The two sets of convective $\alpha$ parameters that we use here are given in
the Table.  The main difference lies in the amount of turbulent viscosity
($\alpha_{\nu}$) which is higher for set~B. Set~A features a small amount of
overshooting ($\alpha_t$) and P\'eclet-correction ($\alpha_r$), which accounts
for the decrease of convective efficiency when radiative losses are important.
Despite the smaller turbulent viscosity in set~A, the models computed with
these parameters show lower luminosity and velocity amplitude by a factor of
two compared to models in set~B.  The smaller $\alpha_{\nu}$ is overcompensated
by the remaining parameters.

Fig.~\ref{fig:obs_alpha} compares the linear and nonlinear phase lags computed
with set~A ( a.) panel) and set~B ( b.) panel) convective parameter sets.  The
Resonance \ML relation and a Galactic metallicity ($Z=0.020$) have been used in
both.  

The shape of the phase lag curves is similar on the two panels, but the range
of the phase lags is smaller for set~B.  The difference has to do with the lower
pulsation amplitudes that are found for set~B.  Another noticeable feature is
the smaller difference between the linear and nonlinear phase lags for panel B.
The underlying cause is again the too low amplitude with the set~B parameters,
because this difference depends on the nonlinear effects, \ie the magnitude of
the amplitude.

The convective parameters affect the modal selection.  Our main concern has
been to reproduce the observed maximum O1 period ($7\dotd 57$ for V440 Per).
We find that the longest period among the nonlinear overtone models is $7\dotd
7$ for the set~A models, and a very small value for the set~B parameter choice
(less than 3 days) that is unacceptable on observational grounds.  The
difference in modal selection manifests itself in the O2 pulsation, too.  For
set~A, there is (pure) second overtone pulsation for mass lower than $3.5$ \Mo.
However the radial second overtone mode is linearly unstable only for masses
below $2.5$ \Mo\ in set~B, and no O2 pulsation was found in the models in the
hydro grid in this case.  This means, that if O2 pulsation exists there, it
would occupy a rather narrow temperature range.

Our experience is that, as long as they are capable of reproducing the observed
mode selection patterns, the turbulent $\alpha$ parameters do not have a major
effect on the phase lags.  

\subsection{Effects of the Metallicity}\label{met_eff}

We have adopted here the following composition pairs (X, Z ) = (0.700, 0.020),
(0.716, 0.010), (0.726, 0.004) as 'canonical' values when computing Galactic,
LMC, and SMC models, respectively.

In Fig.~\ref{fig:metal} we display the linear and nonlinear phase lags for the
LMC ( a.) panel) and SMC ( b.) panel) metallicities.  For short periods the phase 
lag is insensitive to the metallicity, but for high periods (\ie high masses) {\sl
lowering the metallicity causes the theoretical F phase lags to shift upward},
and even causes a reversed sign.  The same trend can be seen for O1 models,
although to a lesser extent. 
We note that the $P_2:P_0 =1:2$ and $P_4:P_1 =1:2$ resonances mentioned in 
Sec.~\ref{sec_res} have very similar effects on the phase lag of the lower 
metallicity model sequences.  Unfortunately no observational extragalactic
$\Delta\Phi_1$ are available to date on which we could test our predictions.

The maximum period of the possible O2 pulsation decreases with decreasing Z,
but observationally, a number of O2 and O1\&O2 double mode Cepheids are
known in the Magellanic Clouds in contrast to the Galaxy.  This does not cause
a problem though, because the evolutionary tracks also shift with $Z$.
Also, our maximum predicted O2 periods are close to maximum O2 period reported 
for O1\&O2 double mode Cepheids in the LMC \citep{s01} and in the SMC \citep{u99}.

So far we have only considered a solar mix where the 'metals' are lumped
collectively into Z.  However, it has been found that deviations from the
standard solar chemical composition \citep{grevesse} play a nonnegligible role
in beat Cepheid period ratios \citep{bs07}.  We have therefore also computed
the linear phase lags for a model sequence with the same opacities as used and
described in \cite{bs07}, namely Fe group element number densities arbitrarily
reduced to 75\% at fixed $Z$. The phase lags are found to be largely
insensitive to the chemical buildup, the difference is largest for high mass
models, being less than 0.01, and in general it is much less.

\vskip 10pt

After having demonstrated that the phase lags are moderately sensitive to
metallicity and that they implicitly depend on the \ML relation and the
turbulent convective parameters, we examine other possible physical and
numerical effects on the Cepheid phase lags.  For these tests we have computed
only {\it linear} models, because the relationship between nonlinear and linear
phase lags is not strongly affected, and this will therefore not significantly
alter our conclusions.

\subsection{Stellar Rotation}

In this section we consider the effect of {\it stellar rotation} because it has 
the effect of rearranging the stellar structure, and can cause slight changes in 
the phase lags. However, since we do not have the 3D code to conduct a proper
calculation of the effects of rotation, we resort to an
expedient subterfuge which we expect to be accurate enough as an order of
magnitude estimate.  As in \cite{bs07} and references therein we incorporate
in our 1D hydrocode a radial pseudo-centrifugal force of the form ($F=\omega^2
r$) which, if anything, is likely to overestimate the effects of rotation.  We
find that even up to 20 $\rm km \ s^{-1}$ rotation, the phase lags are indistinguishable
from the nonrotating ones, except perhaps for the highest masses (>9\Mo), where
the relative difference is still only 0.02.  We note that observed Cepheid
rotation rates ($v_{rot}\thi sin\thi i$) are usually less than 20 $\rm km \ s^{-1}$ \citep{n06}.

\subsection{Numerical Resolution}

Finally, we have tested the effects of the numerical resolution of the models.
In this work we have used 120 zones in our models, which is sufficient to
generate decent light curves.  As the phase lag itself is sensitive to the
structure of the outer stellar layers, different zoning in this region can
potentially influence the lag between the velocity and luminosity maxima.  We
have recomputed some of our Galactic models with 80 and 300 zones.  The linear
phase lags turned out to be rather insensitive to the zoning.  The difference
is the smallest for the low period models, and increases toward higher masses.
The maximum deviation is $\approxlt  0.02$.  Therefore we conclude
that different model zoning has practically no influence on $\Delta \Phi_1$.

\section{Pulsation Amplitudes}\label{sec_amplitudes}

So far we have compared {\it bolometric} theoretical light
curves to observed {\it V magnitudes} when constructing the {\it phase lags}. In
principle, this inconsistency could cause a problem.  Obviously the velocity
curves are free from this caveat.  However, we think that V phase lags are
close to their bolometric counterpart, because the \Teff\ of the Cepheid
variables ensures that the maximum of their spectral energy distribution is
close to the V band.  The $\phi^{mag}_1$ values for V and I, for example, are
indeed found to be very close \citep{ng03}.  For a more detailed comparison of
the V, B, R, I and bolometric light curve parameters we refer to \cite{sm85}.

However, in order to compare our computed amplitudes to the {\it observed
Galactic amplitudes} we need to convert from our bolometric amplitudes to 
V amplitudes: $M_V = M_{bol} + BC$.
For this transformation we follow \cite{kg}, \viz
\begin{eqnarray}
 BC &=& 2.0727 \Delta T  - 8.0634  (\Delta T)^2\\ 
\Delta T &=& \Log \teff - 3.7720
\end{eqnarray}
These fits to the static atmosphere models were derived for Beat Cepheids, but
\citep{bbk} obtained similar formulae for a broader mass range, the difference
being only a small systematic shift between the two ($\Delta \Log L = 0.02$).
Neglecting the {$ log\thi g$} term in the BC formula doesn't affect our 
conclusions. The introduced error is $0\dotm1$ for our lowest and highest 
temperature models, generally being much less than that.

\begin{figure}
\epsscale{1.15}
\ifthenelse{\boolean{color}}
{\plotone{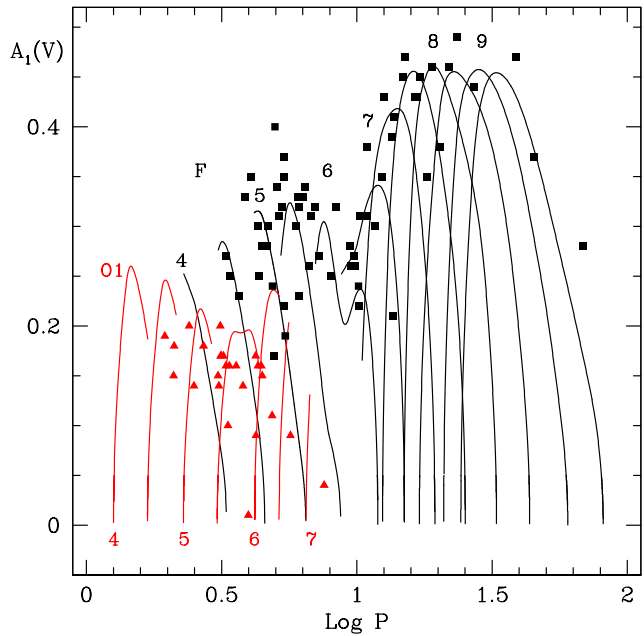}}
{\plotone{fig7.eps}}

\caption{Observed and theoretical values of the $A_1 (V)$ Fourier amplitude,
obtained with the Resonance \ML relation and the $\alpha$ parameter set~A.
The theoretical F models are for our sequences with masses from 4\Mo\ to
9.5\Mo.  The overtone sequences which are located at the left lower side,
correspond to sequences with masses ranging from 4\Mo\ to 7\Mo.  Observed $A_1
(V)$ values from the \cite{ogl00} sample are overlaid for the 
F (squares) and O1 (triangles) Cepheids.}

\label{fig:amplitudes}
\end{figure}

On Fig.~\ref{fig:amplitudes} the $A_1$ Fourier coefficients of the V
light curves are shown for the \cite{ogl00} sample together with our full
amplitude models. The lowest amplitude O1 Cepheid on the figure is $\alpha$ UMi 
\citep{mo00}, included here to complement the observed sample. 
F Cepheids are denoted by squares, overtones by triangles.  The $A_1$ amplitudes 
are plotted for our model sequences with 4\Mo\
to 9.5\Mo\ for F, and with 4 to 7\Mo\ for O1 models, computed by using the
Resonance \ML relation and set~A of convective parameters. Since we have not
computed nonlinear model pulsations at very low amplitude, we have extrapolated
our results to zero amplitude, where applicable, \ie at the F red edge and at
the O1 blue edge.

\vskip 10pt

Fig.~\ref{fig:amplitudes} shows good agreement between the model calculations
and the observations. The only small discrepancy is that the O1
Cepheids do not seem to extend to quite high enough periods, but this partially
is due to the coarse grid we used (100K). Because of this, for the 7.5\Mo\
sequence we do not find stable nonlinear O1 pulsation, while by using a finer
grid we could find a narrow temperature range of overtone pulsation.  

It's important to notice that Fig.~\ref{fig:amplitudes} also appears to show a 
large observational bias toward large amplitude Cepheids, especially among the 
fundamental mode pulsators. This fact has important 
consequences for the observed phase lag distribution. This issue is discussed 
in Sec.~\ref{sec_obs}.

\begin{figure*}
\epsscale{1.15}
\ifthenelse{\boolean{color}}
{\plotone{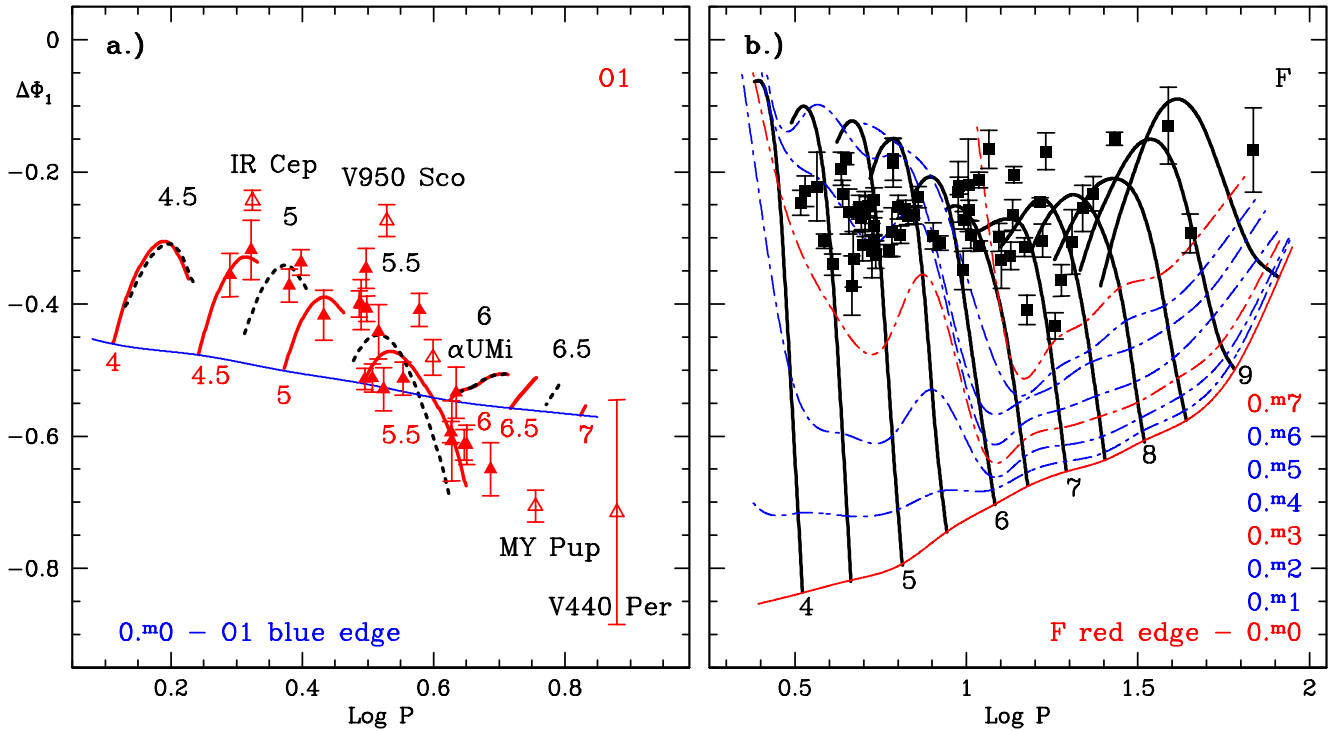}}
{\plotone{fig8.eps}}

\caption{Nonlinear (thick lines) 
phase lags $\Delta \Phi_1$, obtained with he Resonance \ML relation, the $\alpha$
parameter set~A, and a Galactic composition (bottom mass labels). {\bf a.)}: first 
overtone Cepheids, {\bf b.)}: fundamental Cepheids. The observational phase lags of 
the Galactic Cepheids \citep{ogl00} are denoted by triangles and squares, respectively.
Individual O1 stars are discussed in the text. On the left panel, dashed lines
shows models computed with the Girardi \ML relation (top mass labels). On the 
right, V magnitude cutoff lines (dashed lines) help to define the region of normal 
amplitude Cepheids (from $0\dotm 0$ to $0\dotm 7$). The FRE and the OBE are shown 
for reference.}

\label{fig:vmag_cutoff}
\end{figure*}

We now turn to the effects of resonances on our amplitude diagram. 
The action of the $P_2:P_0=1:2$ resonance between
the F mode and the second overtone, which is at the origin of the well known
Hertzsprung progression, is quite visible near $P_0$ = 10\dd\ in the M=6\Mo\
sequence.  It causes a dip in the $A_1$ values. Our models duly follow the 
observed amplitude minimum.

A corresponding 'overtone' resonance between the fourth and first overtone
$P_4:P_1=1:2$ appears at $P_1 = 3\dotd 5-4\dotd 0$ in the O1 sequences of the
M=5.5\Mo\ sequence.  In this case one has to keep in mind that it is easier to
locate the resonance position from the velocity curves than form the light
curves.  \cite{fbk00} found the resonance to be  centered on $4\dotd 2$
(although with a slightly different \ML relation).  The small distortion and
the larger dip between the M=5.5 and M=6\Mo\ overtone amplitude curves
corresponding to this resonance are not seen in the observed sample.  The
likely reason is the small number of observed O1 Cepheids, and the uncertain
mode identification in Galactic Cepheids.

The $P_3:P_0=1:3$ resonance \citep{mb89, am96} has no visible effect,
neither on the computed phase lag nor on the $A_1(V)$ diagrams.  \cite{mb00}
showed that another possible resonance, $P_1:P_0=2:3$ \citep{am96} has no
effect unless it destabilizes the F limit cycle, in which case it causes the
appearance of slightly alternating pulsation cycles (period doubling).  Such
pulsations were reported by \cite{mb01} in radiative Cepheid models, but not
in the sequences of turbulent models that we have computed here.  No
observational evidence has yet shown up to corroborate the existence of
alternating cycles in Cepheids of this period range.

\vskip 10pt

To conclude, we have shown that the amplitudes of our models reproduce the 
observed amplitudes, as well as the resonance characteristics. Furthermore,
we have pointed out a deficiency of observed small amplitude Cepheids on our 
amplitude diagram Fig.~\ref{fig:amplitudes}. These 
features are important regarding the following section, 
in which the observed and theoretical phase lags are compared, and an explanation 
is proposed for the observed phase lag distribution for both F and O1 Cepheids.

\section{Comparison With Observation} \label{sec_obs}

\subsection{The fundamental mode phase lag}

At this time only phase lags from Galactic Cepheids have been reported in the
literature \citep{ogl00,mo00}.  In Fig.~\ref{fig:obs_alpha}a we have
plotted these observational Galactic phase lags.  The F Cepheids and O1
Cepheids denoted by squares and triangles, respectively.  They are superposed on
the linear and nonlinear phase lags that we have computed with our 'standard'
reference parameters in this paper, namely the convective parameter set~A ( a.) panel), 
the Resonance \ML relation and a Galactic metallicity ($Z=0.020$).

In Figs.~\ref{fig:linNL}a and \ref{fig:obs_alpha} it is seen that our model
sequences extend over a much broader range of periods than the observational
data.  The simple reason is that our sequences are not constrained by the
evolutionary tracks (except for the \ML relation), and thus they cover a much
larger range of stellar parameter values ($L$, $M$, \Teff).

In addition, the observed stars strikingly occupy only the upper section of the
region in the $\Delta \Phi_1$ \vs Log $P$ diagram that is permitted by the
modeling.  This region corresponds to the inside region of the IS;\thi in other
words the observed F sample seems to avoid the vicinity of the F red edge
(Fig.~\ref{fig:linNL}a). We argue that this arises from an observational 
amplitude bias. Fig.~\ref{fig:amplitudes} suggests that the low amplitude Cepheids 
are missing from the 
observed sample, most likely because of observational selection effects. We note
though that there is also a theoretical reason for the absence of low amplitude
Cepheids \citep{bk02}.  In all our Log P - $\Delta \Phi_1$ figures,
Figs.~\ref{fig:linNL}a and \ref{fig:MLrel} the bottoms of the downward legs of
the fundamental $\Delta \Phi_1$ curves correspond to the FRE and thus to models
with pulsation amplitudes that vanish as the FRE is approached.  In addition,
towards the red edge the phase lags of the F Cepheid models are extremely
sensitive to temperature as witnessed from the spacing of the dots in
Fig.~\ref{fig:linNL}a. 

Fig.~\ref{fig:vmag_cutoff}b juxtaposes the theoretical nonlinear
F phase lags and the \cite{ogl00} sample.  As discussed in \S4 the F models
sequences terminate at finite amplitude at the nonlinear fundamental blue edge
(NFBE), because the model sequence switches either to double mode or to O1
single mode pulsation at that point.

Overall the agreement of the model phase lags with the observational ones is
good (differences of 0.1 rad translate to only about 6$\degr$).  However, a few
systematic discrepancies stand out.  For example, the observational F Cepheid
phase lags fall slightly above the theoretical ones in the middle range of
periods.  But, from a comparison of Fig.~\ref{fig:obs_alpha}a and
Fig.~\ref{fig:metal} we see that this discrepancy would be mitigated or even
removed if we decreased the metal content that we have assumed for the Galactic
Cepheid models, in line with a recent revision of the solar metal abundances
\citep{asplund}.

In Fig.~\ref{fig:vmag_cutoff}b lines of pulsation equi-amplitude
(cutoffs) are overplotted from $0\dotm 0$ at the FRE to $0\dotm 7$.
To avoid confusion we caution that these lines refer {\sl only} to the rising
parts of the $\Delta\Phi_1$ curves near the FRE.\thi (There may be multiple
equi-amplitude lines because the pulsation amplitudes are not just a function
of period and $\Delta\Phi_1$;\thi the same amplitude can occur more than once
along the same arch, but generally not on the same line.)

Based on Fig.~\ref{fig:vmag_cutoff}b we can see that the lower limit of the 
observed amplitude distribution is well represented by the $0\dotm 7$ and $0\dotm 3$ 
cutoff lines for $P\approxgt$10\dd and $P\approxlt$10\dd, respectively. This dichotomy and 
the sharp change in the equi-amplitude lines on the $\Delta \Phi_1$~--~log~P
plane around 10 days has its origin in the $P_2:P_0=1:2$ resonance discussed
above. Fig.~\ref{fig:vmag_cutoff}b together with Fig.~\ref{fig:amplitudes} 
confirm our hypothesis that observational selection effects make the observed 
sample gravitate toward the high amplitude region.

\subsection{The first overtone phase lag}

On Fig.~\ref{fig:vmag_cutoff}a we display the region of O1 models separately, to 
avoid overcrowding. Five individual stars that are mentioned 
below are labeled and plotted with open symbols. As discussed in \S4 the O1 
models sequences terminate at finite amplitude at the NORE, because the model 
sequences continue as F single mode pulsation to the right of that point.  
The $P_4:P_1=1:2$ resonance is seen to have a remarkable effect on the M=5.5\Mo\ 
and M=6.0\Mo\ phase lags.  As is seen most sharply in the 5.5\Mo\ sequence the 
resonance causes $\Delta\Phi_1$ to drop sharply below its value at the blue 
edge as the nonlinear overtone red edge (NORE) is reached.

In Fig.~\ref{fig:vmag_cutoff}a we have also shown the phase lags
of 4.5, 5.0, 5.5, 6.0 and 6.5\Mo\ sequences computed with the Girardi \ML relations.
They are shown as dashed lines.  (The 6\Mo\ curves are almost indistinguishable
for the two different \ML relations).  Interestingly we find that while the low
mass curve is shifted to considerably smaller periods, the envelope of the
accessible $\Delta\Phi_1$ curves is affected only slightly. (The reason for
the shift to lower periods is of course consistent with Fig.~\ref{fig:masslum} where
the Girardi \ML relations dip strongly below the Resonance \ML relations.

In Fig.~\ref{fig:vmag_cutoff}a the O1 observational points seem to fall a little higher
than the ones computed for Galactic models.  Again, reducing the metallicity
would at least partially mitigate the difference.  Most of the discrepancy
occurs because of 3 really egregious stars, two with 
$\Phi_1>-0.3$ (IR Cep and V950 Sco), and the long period V440 Per with
$\Phi_1<-0.7$.  For the first two, strong evidence has been given on the basis
of the Fourier decomposition parameters of the light and radial velocity curves
that they are O1 pulsators (Moskalik, private communication).  However,
according to their location in the phase lag diagram, Fig.~\ref{fig:obs_alpha}a, at
least, they would quite naturally fit in the F, rather than in the O1 group.
Alternatively, the disagreement for IR Cep would also disappear if its
metallicity was somewhat below the average Galactic metallicity, although this
would not apply as well for the longer period V950 Sco.

V440 Per has a low amplitude of $A_1$=0.095, and an enormous error bar for
$\Delta\Phi_1$ \citep{ogl00}.  It is classified as the longest period
Galactic O1 Cepheid.  One notes that the O1 Cepheids MY Pup and its immediate
neighbors to the left fall right in the middle of the resonance region.  The
employed \ML\ relation has an effect on the location of this resonance, and
this in turn affects the results for $\Delta\Phi_1$.  A comparison between the
phase lags in Fig.~\ref{fig:vmag_cutoff}a with the Girardi and the Resonance \ML
relations suggests that an \ML\ relation that lies a little higher than the
Resonance \ML\ in Fig.~\ref{fig:masslum} in the 5.5 to 6.0 \Mo\ range should push
the computed $\Delta\Phi_1$ a little further to the right and improve the
agreement for the stars with $\Log P_1>0.6$, and in particular for MY Pup.
However, it is unlikely that this would solve the problem with V440 Per, but we
note that the phase lag of this star is compatible with the theoretical ones
for small amplitude F Cepheids as seen in Fig.~\ref{fig:obs_alpha}a, suggesting that
perhaps it is a small amplitude F Cepheid rather than an overtone Cepheid.

$\alpha$ UMi is a well-known Cepheid, most probably pulsating in the
first overtone mode. This star has been showing dramatic amplitude changes in 
the last decades \citep{t05}. 
Currently being a low amplitude pulsator with $A_V = 0\dotm015$ \citep{f95}, it 
resides relatively close to our blue edge on the Log P - $\Delta \Phi_1$ plot\ 
\citep{mo00}. However this agreement should be taken with a grain of salt, 
because this star can be in the phase of crossing the instability strip for 
the first time \citep{t05} thus bearing a different \ML relation than most of 
the Cepheids (see \eg Fig.~2 of \cite{bs07}). If this is the case, then 
evaluating it's position on Fig.~\ref{fig:vmag_cutoff}a would require more 
computational work with the appropriate \ML relation, which is out of scope of 
this paper.
 
Finally we note that on a large scale the O1 phase lag distribution appears
essentially linear \citep{ogl00}.  If the three egregious stars are removed
from the O1 Cepheids then not only does the observational phase lag distribution appears 
curved, but that our results closely match its general shape (albeit being a little lower).
We have also shown that the curvature of the distribution has its origin in the
$P_1$ : $P_0$ = 2:3 resonance.

\subsection{The phase lag of other radially pulsating Cepheids}

Recently ultra-low amplitude (ULA) Cepheids have been discovered in the LMC
\citep{bwk05}.  Based on this work we predict that ULA Cepheids pulsating in
the F mode and close to the fundamental red edge, would feature a greater
(negative) phase lag, and would fall below the current observed range of the 
regular amplitude Cepheids. Also, ULA O1 Cepheids with periods far from the resonance 
center  would be close to the lower envelope of the predicted phase lag
distribution. 

Our Fig.~\ref{fig:obs_alpha} offers a method for radial mode identification for
Cepheids that is additional and complementary to the usual relative Fourier
decomposition coefficients ($R_{21}, \Phi_{21}$, etc).  We see that it offers 
not only a criterion for the F and O1 Cepheids, but also for the elusive
Galactic O2 Cepheids.  There clearly is an observational bias against low
period - low amplitude Cepheids.  However, considering the existence of a
Galactic O1\&O2 beat Cepheid (CO Aur) it is quite likely that pure O2 Cepheids 
exist as well. The phase lag has the potential of sorting the elusive Galactic O2
Cepheids from O1 Cepheids.

\vskip 10pt

Observed and theoretical Fourier parameters of Cepheid light curves have been 
often investigated, see \eg \cite{sk95} for long period F Cepheid Fourier 
decomposition, \cite{sd83} for F phase lags. Other previous theoretical works on 
O1 Cepheid pulsations \citep{aa95} and on O2 Cepheid pulsations \citep{ak97} 
Cepheids also discussed the phase lags. 
However, these authors ignored convection and produced purely {\it radiative}
models, an approximation which makes sense only near the blue edge.  The
resulting large range of F and O1 phase lags led \cite{aa95} to conclude that
$\Delta \Phi_1$ is not a useful discriminator for the pulsational mode.  One
problem may be that {\it radiative} models have excessively large amplitudes. 
Because the nonlinear phase lags depend largely on the
amplitude, as we pointed out in Sec.~\ref{alphas} we think that to address the
phase lag problem correctly, convection can not be left out from the models.

\section{Conclusion}

In this paper we have reexamined the classical Cepheid phase lags, $\Delta
\Phi_1= \phi^{V_r}_1 - \phi^{mag}_1$, with our hydrocode which includes a
time-dependent mixing length model for turbulent convection. 
 
As expected, the linear and nonlinear (full amplitude) phase lags approach each 
other when the pulsation amplitude is low, as for example near the fundamental 
red edge of the instability strip. The nonlinear
phase lags generally fall below the linear ones except for high period models
where they lie above.  The differences between the linear and 
nonlinear phase lags (\DDP) are generally at most 0.20  
for our Galactic models with the best convective model parameters (set~A).  
In case of lower metallicities, the \DDP\ can be slightly larger.

In the linear regime we find that high mass models of low metallicity 
can have positive $\Delta \Phi_1$.  However, 
for Galactic composition {\sl the nonlinear $\Delta \Phi_1$ are always 
found to be negative}, \ie the radial velocity always lags the luminosity slightly. 

There is a small metallicity effect:  
Lower metallicity models pulsating in the fundamental mode show phase lags that are 
slightly higher than the Galactic values.  There are however no sufficiently
good observations to confirm this. 

The convective $\alpha$ parameters have a strong influence on the modal selection in 
the instability strip, as well as on the pulsation amplitude, which is then reflected 
in the Period -- $\Delta \Phi_1$ diagram.  Similarly, the \ML\ relations have only an 
indirect effect on the phase lags through a modification of the mass--period relation.

We obtain a good general agreement between the observed Galactic Cepheid phase
lag data of \cite{ogl00} and our models.  For the F Cepheids this agreement can be 
further improved when the observed amplitude distribution is taken 
into account.  By applying cutoff magnitude (equi-amplitude) lines we can restrict 
the model phase lags region to the approximate observed range [$-0.2;-0.4$].
An almost perfect agreement can be achieved by lowering the assumed metallicity of 
$Z=0.020$ in accordance with the suggestion of \cite{asplund}.

Our computed O1 phase lags seem to fall on a steeper relation than the
observations, but this discrepancy would be mitigated or removed if we could
reclassify two stars, IR Cep and V950 Sco as F Cepheids (even though Fourier
decomposition parameters of the light and radial velocity curves indicate
otherwise).  Furthermore, the phase lag of the outlier, V440 Per, which is
normally classified as an overtone, would be compatible with the theoretical
phase lag of a low amplitude F Cepheid.  With the omission of these three stars
our results match the shape of the observational distribution of phase lags.

We have shown that resonances have an impact on the phase lag.
The $P_2:P_0=1:2$ resonance which is present in the fundamental mode Cepheids at around 
10\dd\ period  and
causes the well-known Hertzsprung-progression, has a small effect on the minimum
in the nonlinear phase lags (Fig.~\ref{fig:linNL}a and Fig.~\ref{fig:obs_alpha}a). 
As for the O1 models, the $P_4:P_1=1:2$ resonance generates an abrupt widening
of the  'either 
F or O1' region at around the resonance center (Szab\'o \etal 2007, in preparation).
In this region $\Delta \Phi_1$ is 
more negative than for models computed far from the resonance, and this effect seems 
to have great influence on the observed O1 phase lag distribution.

We confirm the finding of \cite{ogl00} and \cite{simon84} that the $\Delta
\Phi_1$ is a good indicator of the pulsational mode that is independent of and
supplemental to the usually used relative Fourier decomposition coefficients
($R_{21}, \Phi_{21}$, etc).  In principle, it is also a good criterion for
identifying second overtone pulsators.  However, no O2 Cepheids have been
detected in the Galaxy, but there are a number of candidates in the Magellanic
Clouds.  For an independent confirmation of their status as O2 on the basis of
the phase lag it would be necessary to acquire radial velocity data on these
objects.

Ultra-low amplitude Cepheids should be easily discernible based on their 
phase lags. Our work predicts that ULA Cepheids pulsating in the
fundamental mode and close to the fundamental red edge, would feature a greater
phase difference, and would lie below the observed range [$-0.2;-0.4$] of the
large-amplitude Cepheids. ULA first overtone Cepheids (at the blue edge) 
would be similarly separable, except in the vicinity of the 'overtone' resonance.
Therefore obtaining radial velocity  data and existing light curve parameters could 
further prove their nature, \ie radial
pulsation. Unfortunately, getting \RV\ curve for a small amplitude Cepheid in
the MCs remains a huge technological challenge.

Our survey has produced concurrently a rather complete theoretical overview of
the nonlinear Cepheid instability strip, which delineates the regions of F, O1,
O2 pulsation, of F\&O1 and O1\&O2 beat pulsation, and regimes of hysteresis,
such as the 'either F or O1' regions and 'either O1 or O2' regions.  These
findings will be presented in a separate paper \citep{sb07}.

\acknowledgments

It is a pleasure to thank Peter Wood and Zolt\'an Koll\'ath for valuable
discussions, as well as Pawel Moskalik for making the observed data available
and his comments on the nature of IR Cep and V950 Sco.  This work has been
supported by NSF (AST03-07281 and OISE04-17772) at UF.  RSz acknowledges the
support of a Hungarian E\"otv\"os Fellowship.  We are grateful to the the
Hungarian NIIF Supercomputing Facility and to the UF High-Performance Computing
Center for providing computational resources and support without which we would
not have made such extended survey of full amplitude models.


\end{document}